# Ferromagnetism in Cr-Doped GaN: A First Principles Calculation

G. P. Das,[1,2] B. K. Rao,[1] and P. Jena[1]
[1]Department of Physics, Virginia Commonwealth University
Richmond, VA 23284-2000
[2]Bhabha Atomic Research Centre, TPPED, Physics Group
Mumbai 400 085, India

## Abstract

Otherwise antiferromagnetic chromium is shown to couple ferromagnetically when doped into GaN irrespective of whether the host is a crystal or a cluster. The results on the doped clusters and crystals are obtained from density functional theory based molecular orbital theory and linearized muffin tin orbital tight binding supercell band structure method respectively. The calculated half metallic behavior of Cr-doped GaN crystal combined with the observed room temperature Curie point make it an ideal candidate for spintronics applications.

The interplay between the charge carriers in a semiconductor and the electron spins of a ferromagnetic metal doped into the semiconductor can be utilized for many magneto/spin-electronic devices. Interest in this family of dilute magnetic semiconductors (DMS) started with the discovery of ferromagnetism in ~5% Mn doped InAs and GaAs with a Curie point of 110K[1] and the subsequent theoretical prediction[2,3] that the Curie point in Mn doped GaN could be higher than the room temperature. Amongst the III-V semiconductors, GaN, with a wide band gap of 3.5eV, is extensively studied because of its promising applications to short wavelength opto-electronic devices.[4] The possibility of making it ferromagnetic by injecting into it spin-polarized electrons from a 3d transition metal further increases its potential for spintronics applications. In particular, Mn doped GaN has been found to be ferromagnetic irrespective of the method of growth, although as far as the magnitude of the Curie temperature is concerned there is controversy[5] which is yet unresolved.

The recent discovery[6] of room temperature ferromagnetism of Cr-doped GaN single crystals has stirred further interest in the DMS systems. This discovery is significant in many ways: (1) Although both Mn and Cr are antiferromagnetic in the bulk phase, the cohesive energies of Cr is significantly larger than that of Mn. Similarly while the $Cr_2$-dimer is strongly bound $Mn_2$–dimer is very weakly bonded.[7] Furthermore, the coupling between the Mn spins in $Mn_2$-dimer is ferromagnetic while that between Cr spins in $Cr_2$-dimer is antiferromagnetic. (2) While Mn-doped GaN samples are grown in the form of thin films,[8] Cr-doped GaN has been successfully grown[6] in the form of bulk single crystals using a sodium flux growth method. The X-ray diffraction pattern of bulk GaN:Cr single crystal clearly shows a wurtzite structure with c = 5.191 Å which is marginally larger than that of pure GaN (c = 5.185 Å). This is attributed to the larger Cr atomic radius (1.40Å) than Ga (1.26 Å), indicating that Cr indeed substitutes the Ga site. (3) The ferro- to para-magnetic transition temperature ($T_c$) is observed to be around 280K from SQUID magnetization measurement as well as from temperature dependence of electrical resistivity that shows a change in slope at ~280K. A higher $T_c$ ($\geq$ 400K) has also been reported for thin films of GaN:Cr grown on sapphire substrate by ECR-MBE technique.[9] However, in case of thin films, the contribution of substrates plays a crucial role.

A magnetic impurity such as Mn which is in $3d^5$ $S=5/2$ configuration, contributes spins as well as holes (p-type doping) to the III-V semiconductors such as GaAs or GaN, which therefore experiences carrier-controlled ferromagnetism. Dietl *et al.*[2] used mean field Zener model of ferromagnetism to predict how incorporation of Mn ions induces ferromagnetic behavior via hole-mediated exchange interaction (RKKY type). Litvinov and Dugaev[5] subsequently proposed the indirect exchange interaction caused by virtual electron excitations from magnetic impurity acceptor levels to the valence band. There are very few first principles investigations[10,11] that have been attempted to understand the origin of the ferromagnetic coupling in Mn-doped GaN systems and to explain the stability the system as a function of Mn concentration. To the best of our knowledge there has been no first principles investigation reported so far on the ferromagnetism of Cr doped GaN.

In this work we report the first theoretical study of the electronic structure, energetics and magnetism of Cr doped GaN in cluster as well as bulk crystalline environment. These results are based on density functional theory with generalized gradient approximation of exchange-correlation potential. In order to study the effect of Cr impurity on GaN crystal, we have followed the supercell band structure approach with the cation (Ga) site selectively replaced by Cr atoms. Alternatively the DMS can be considered as a substitutionally disordered alloy which can be treated[12] using coherent potential approximation. The latest experimental work on GaN:Cr single crystals clearly reveal[6] diffraction pattern corresponding to hexagonal wurtzite GaN structure with Cr substituting Ga sites. Accordingly, all our self-consistent band structure calculations in this work have been performed with supercells that are 8 times larger than the wurtzite GaN unit cell. We have used experimental values of lattice parameters eg. a = 3.189 Å, c = 5.191 Å. This c/a ratio 1.628 is slightly larger than that of pure GaN, and is due to the larger Cr atomic radius (1.40 Å) compared to that of Ga (1.26 Å). The corresponding Ga-N bond length is 1.947 Å. In the wurtzite structure, the lattice vectors are (in units of a) ($\sqrt{3}/2,-1/2,0$), (0,1,0) and (0,0,c/a) while the unit cell contains two cations at (0,0,0) and (2/3,1/3,1/2), and two anions at (0,0,3/8) and (2/3,1/3,7/8). In the supercell, the lattice vectors have been doubled along all the 3 directions, thereby accommodating 16 Ga and 16 N atoms. For studying Cr-doped GaN, we replaced a pair of neighboring Ga atoms by Cr atoms so that the super cell formula unit becomes $Ga_{14}Cr_2N_{16}$. The reason for substituting two Ga atoms by Cr is to allow the Cr spins the freedom to couple ferromagnetically or antiferromagnetically. We should point out that the 32 atom super cell is one of the smallest super cell that ensures separation between the impurities in neighboring super cells by at least a few times the Ga-N bond length.

We have used self-consistent tight-binding linear muffin-tin orbital (TB-LMTO) method[13] with the Atomic Sphere Approximation (ASA) and incorporated the "Combined Correction" term that accounts for the non-ASA contribution to the overlap matrix.[14] Knowing the ASA potential, one can calculate the ground state charge density and hence the ASA total energy. For exchange-correlation potential we have used the local spin-density approximation to DFT, along with the generalized gradient correction as per the original Perdew-Wang formulation.[15] All our calculations have been performed scalar relativistically i.e. without spin-orbit interaction which is not significant in case of GaN. Spin polarized bands have been calculated with minimal basis set consisting of s-, p-, and d-orbitals ($\ell = 2$) for Ga, Cr and N, with N-d orbitals downfolded. The need for treating the localized semicore 3d states of Ga as band states has already been emphasized by other workers,[16,17] and accordingly we have included the fully occupied 3d-states of Ga as relaxed band states in our self-consistent iterations. Apart from the valence states of Ga, Cr and N, the core orbitals were kept frozen to their isolated atomic form. In order to fix the magnitudes of sphere radii of the individual atoms, we have used the so-called Hartree potential plot prescription.[18] This yields Ga and N atomic sphere radii to be 1.227Å and 1.015Å, which are roughly proportional to the corresponding covalent radii of 1.62Å and 1.26Å of Ga and N respectively. And for Cr, we have used the same atomic sphere radius as that of Ga. Since the wurtzite GaN is an open structure, which is typical of any tetrahedrally bonded semiconductor, we have introduced two different types of empty spheres (2 of each type) at the high symmetry positions consistent with the P6$_3$mc space group. Thus the wurtzite unit cell is divided into a total of eight overlapping atomic spheres (4 real atoms and 4 empty spheres). This translates to a 64 sphere supercell, i.e. 32 real atoms and 32 empty spheres. In all our super cell calculations, we have used (6,6,4) **k**-mesh which corresponds to 144 **k**-points in the irreducible wedge of the simple cubic BZ. Brillouin Zone (BZ) integration has been performed using the improved tetrahedron method[19] that avoids mis-weighting and corrects errors due to the linear approximation of the bands inside each tetrahedron.

Many of the salient features of the electronic structure of GaN:Cr system can be seen from the total DOS per unit super-cell (Fig. 1) and the partial DOS of the Cr impurity atom (Fig. 2). First of all we observe a half-metallic behavior in the sense that the Fermi level state density is finite for the majority spin and zero for the minority spin. The Cr d-band with a 3-peaked structure and a width of ~2.5eV falls within the LDA gap of GaN. Both the doubly degenerate $e_g$ bands and the triply degenerate $t_{2g}$ bands of Cr are strongly spin-split by ~1.8eV. While the majority states of Cr lie within the GaN bandgap, as can be seen from the $\ell$-projected fat bands (Fig. 3), the minority states merge with the conduction band states. The bonding peak for Cr majority spin lie ~1.5eV above the valence band top, which matches with the estimate given by van Schilfgaarde and Mryasov.[20] This implies that Cr acts as an acceptor.

Our calculations show that $Ga_{14}Cr_2N_{16}$ is having a ferromagnetic ground state with a localized magnetic moment of 2.69 $\mu_B$ on the Cr atoms. The nearest neighbor host atoms are weakly polarized (induced moments of +0.025 $\mu_B$ on n.n. Ga sites and –0.025 $\mu_B$ on n.n. N sites). Similar calculations that we had performed earlier[10] on $Ga_{14}Mn_2N_{16}$ supercell had shown a magnetic moment of 3.5 $\mu_B$ on the Mn atom. Thus doping of GaN by Mn or Cr, both of which are nearly half-filled bands leads to ferromagnetism. This result agrees with the conclusion derived from KKR-CPA calculations,[12] where the authors reported the trend in the magnetic states when different 3d transition metals are doped into III-V as well as II-VI semiconductors. They also had arrived at the conclusion that Cr- and Mn-doped GaN DMS's are both ferromagnetic. It is to be noted that increase of Cr-Cr separation in the supercell, or incorporating Cr in the interstitial site leads to drastic reduction in the magnetic coupling.

In order to see if the ferromagnetic coupling between Cr spins is governed by its local environment, we have studied an extreme situation by considering atomic clusters of $(GaN)_xCr_2$ ($x \leq 3$). In this size range, almost all the Ga and N atoms are surface atoms and consequently have dangling bonds. Interaction of the Cr atoms with the Ga and N atoms not only can alter the properties of $(GaN)_x$ clusters, but also can influence their magnetic coupling with each other. We have studied this problem by using the linear combination of atomic orbitals-molecular orbital (LCAO-MO) method and gradient corrected density functional theory. The equilibrium geometries of $(GaN)_x$ and $(GaN)_xCr_2$ clusters are computed by optimizing their structures with respect to different arrangements as well as spin configurations. Since the clusters have finite number of electrons occupying discrete molecular energy levels, their preferred spin multiplicity (M = 2S+1) yields the number of unpaired electrons and hence the total magnetic moment. An analysis of the spin density distribution and the Mulliken population yields the directions of the spin moments (up or down) and hence the magnetic coupling.

We have used frozen core LANL2DZ basis sets for the atomic orbitals and the calculations were performed using the Gaussian 98 code.[21] We have used the Becke, Perdew, Wang (commonly referred to as BPW91) prescription for the GGA form of the exchange correlation potentials. The threshold of the root mean square forces for geometry optimization were set at 0.0003 a.u./Bohr. Calculations were carried out for various spin multiplicities of each cluster. Here we only list the results for the ground states. The remaining results will be published elsewhere.

In Fig. 4 we present the equilibrium geometries of $(GaN)_x$ and $(GaN)_xCr_2$ (x ≤ 3) clusters. We note that the structures of $(GaN)_x$ (x ≤ 3) clusters are planar with bonds only between Ga and N atoms. The GaN bond length hardly changes with cluster size. This situation changes substantially when Cr atoms are added. The GaN bonds break in $(GaN)_x$ (x ≤ 2) clusters but reappear in $(GaN)_3$ although in a stretched form. On the other hand, the Cr-N bond emerges and the corresponding bond length remains relatively unchanged throughout. The energy gains, Δ, in adding a $Cr_2$ dimer to an existing $(GaN)_x$ cluster, namely, $\Delta = -[E(GaN)_xCr_2 - E(GaN)_x - E(Cr_2)]$ are 5.43, 4.60, and 6.80 eV respectively for x = 1, 2, 3. The corresponding values of Cr-Cr distances are 3.29 Å, 2.68 Å, 274 Å respectively for x = 1, 2, 3. In this context it is worth mentioning that the calculated binding energies of CrN and GaN are respectively 2.06 eV and 2.16 eV. Consequently it is energetically possible to replace Ga by Cr in GaN and the success of synthesizing Cr-doped GaN in single crystals has a molecular origin.

We next discuss the magnetic configurations of these clusters. The total magnetic moments of $(GaN)_x$ clusters are 2, 2, and 0 $\mu_B$ respectively for x = 1, 2, 3, i.e. GaN clusters containing as few as three dimers cease to be magnetic. However, as $Cr_2$ is added, the magnetic nature changes completely. The total magnetic moment of $(GaN)_1Cr_2$ is 8 $\mu_B$ out of which 0.07 $\mu_B$, -0.64 $\mu_B$, and 4.28 $\mu_B$ moments are residing at Ga, N, and Cr sites respectively. Note that the coupling between two Cr atoms is ferromagnetic while that between Cr and N is antiferromagnetic. Similar results are obtained for $(GaN)_2Cr_2$ and $(GaN)_3Cr_2$ clusters although the magnitudes of the moments vary with size. In $(GaN)_2Cr_2$, the moments at Ga, N, and Cr sites are respectively –0.32, -0.32, and 1.64 $\mu_B$. The total magnetic moment of $(GaN)_2Cr_2$ is significantly reduced from that in $(GaN)Cr_2$, namely to 2 $\mu_B$ although Cr atoms remain ferromagnetically coupled. The total magnetic moment of $(GaN)_3Cr_2$ rises sharply to 10 $\mu_B$ of which 0.66 $\mu_B$, -0.04 $\mu_B$, and 4.07 $\mu_B$ moments reside at Ga, N, and Cr sites. Once again the coupling between the Cr atoms is ferromagnetic. It is important to note that the coupling between the Cr and the N atoms which are strongly bonded to each remain antiferromagnetic irrespective of whether Cr is doped into clusters or crystal. This provides some insight into the origin of the ferromagnetic coupling between doped Cr atoms – namely that it is an indirect exchange mechanism mediated by nitrogen.

At this point, we should note that the magnetic coupling of Cr atoms in the presence of $N_2$ was studied some time ago by Weber *et al.*[23] Here the authors found that as the $N_2$ molecule approached a $Cr_2$ dimer, the nitrogen-nitrogen bond got cleaved at a distance of 0.6 Å and the antiferromagnetic coupling between the Cr atoms in the $Cr_2$ dimer changed to ferromagnetic coupling with a net moment of 6 $\mu_B$. The possibility that chemical reaction can cause magnetic transition allows new possibilities for designing magnetic devices.

In conclusion we have shown through first principles calculations that the Cr atoms couple ferromagnetically when doped into GaN whether the host is a cluster or a crystal. This coupling is mediated by nitrogen through an indirect exchange mechanism. The magnetic moment per Cr atom is about 4 $\mu_B$ in clusters while it is about 2.69 $\mu_B$ in bulk GaN crystal. That the coupling is ferromagnetic irrespective of the host structure allows great flexibility in synthesizing DMS systems involving Cr-doped GaN.

This work was supported in part by a grant from the Department of Energy (DEFG02-96ER45579).

**Figure Captions**

Fig. 1. Total density of states of $(Ga_{14}Cr_2)N_{16}$ supercell for majority spin (top) and minority spin (bottom).

Fig. 2. Partial DOS of Cr in $(Ga_{14}Cr_2)N_{16}$ supercell majority spin (top) and minority spin (bottom).

Fig. 3. Cr-projected fat-bands for majority spin (a) $e_g$ projected (b) $t_{2g}$ projected.

Fig. 4. Ground state cluster geometries of $(GaN)_x$ (left panel) and $(GaN)_xCr_2$ (right panel).

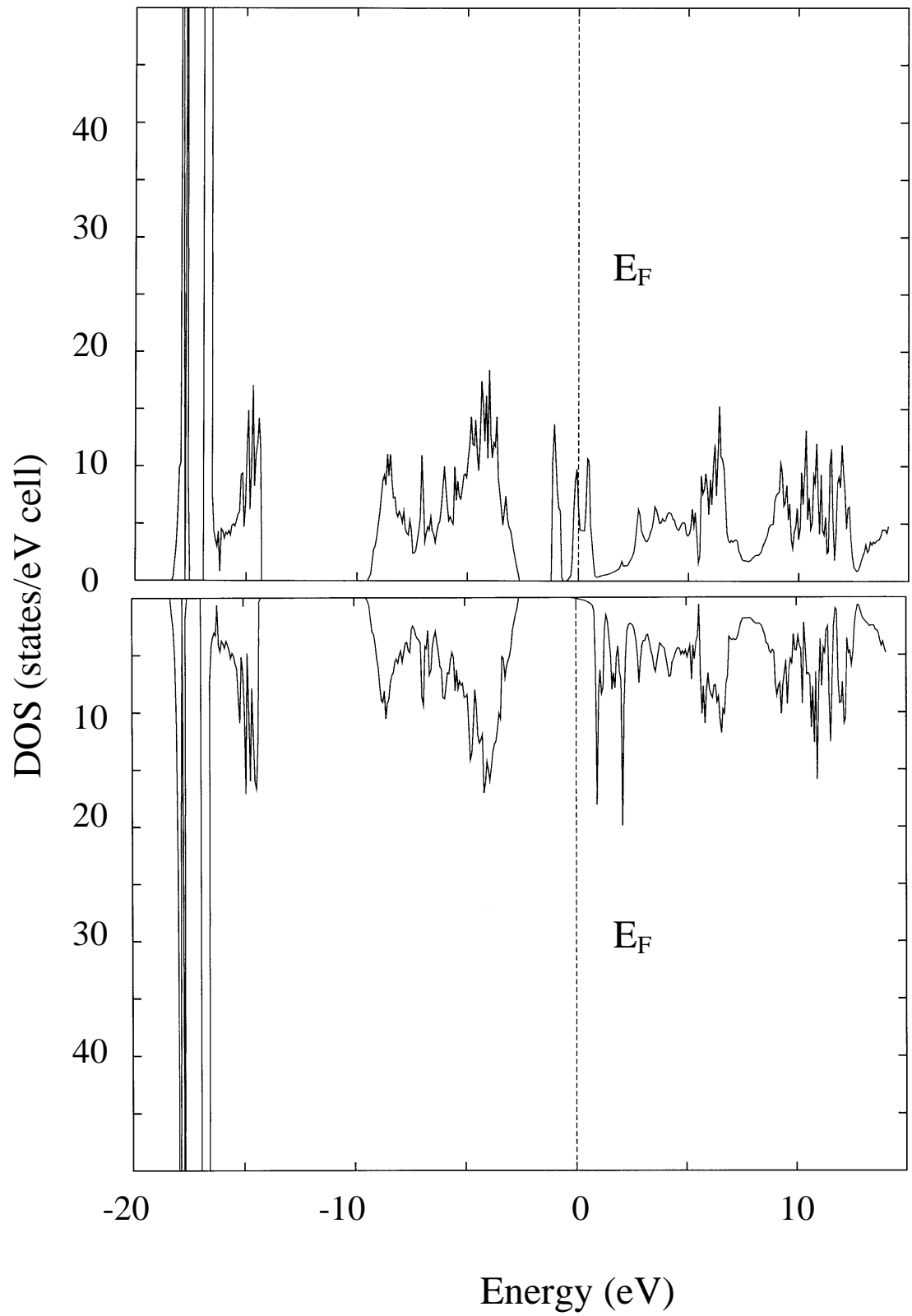

Fig. 1

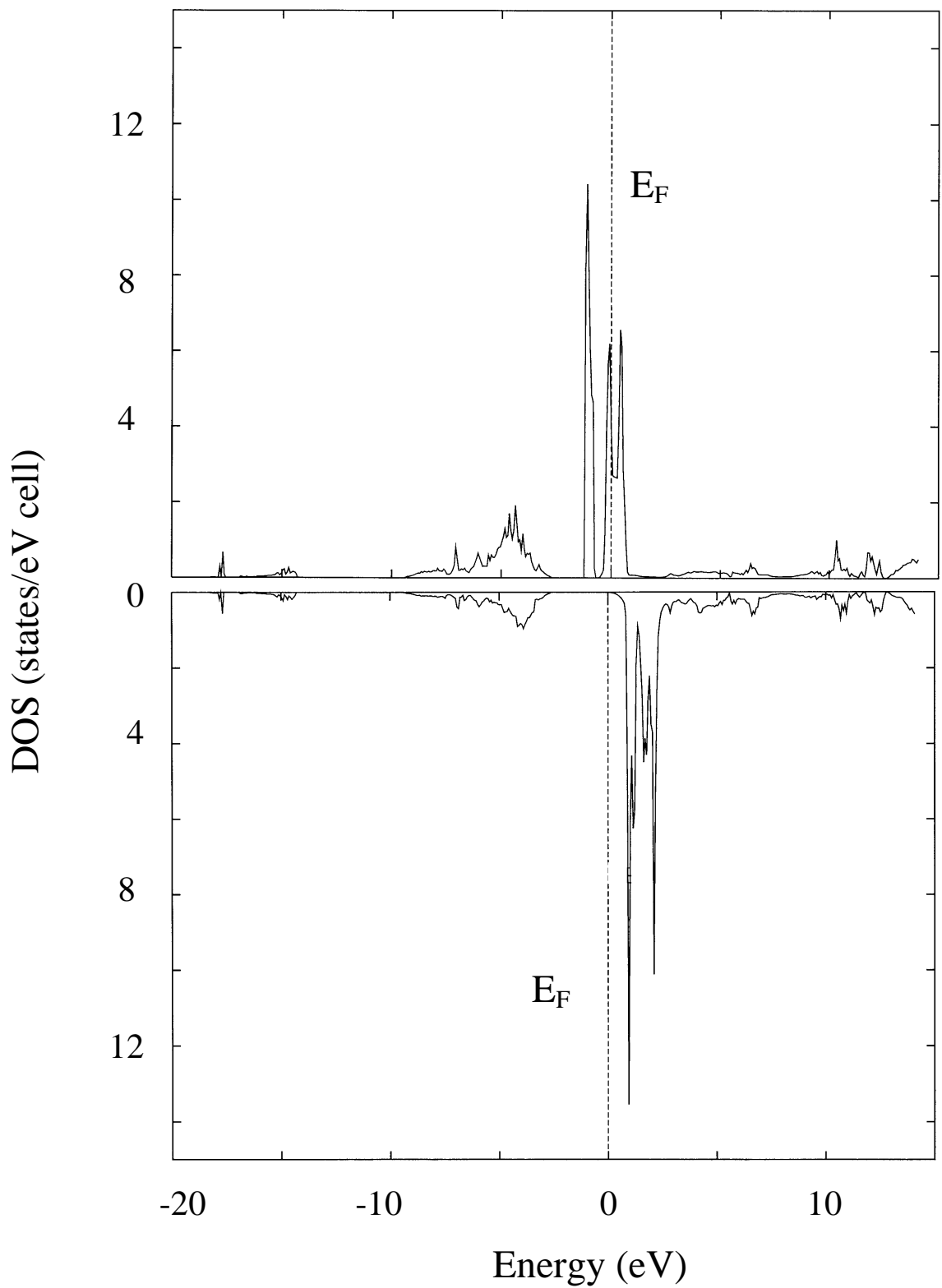

Fig. 2

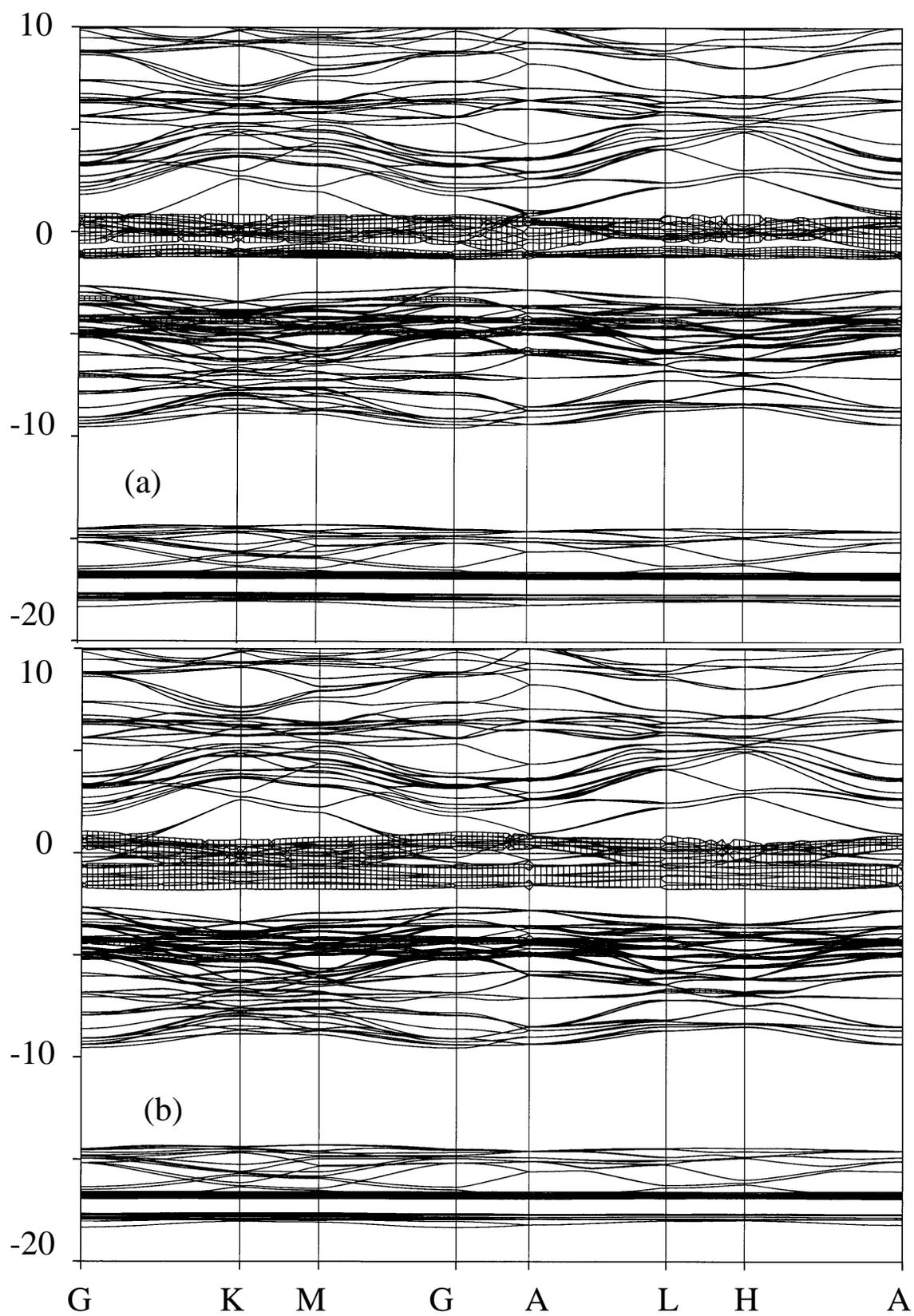

Fig. 3

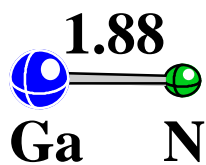

(a) Ga—N 1.88  2μ_B

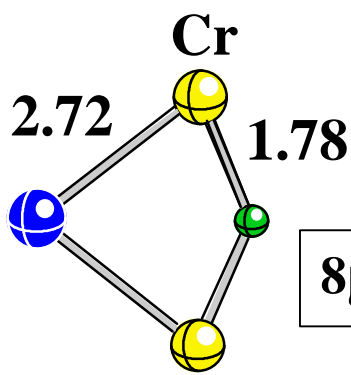

(b) Cr 2.72 1.78  8μ_B

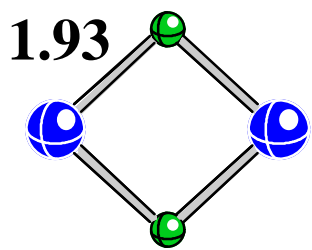

(c) 1.93  2μ_B

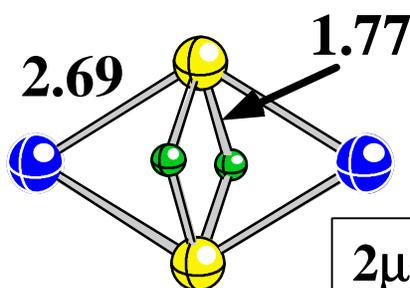

(d) 2.69 1.77  2μ_B

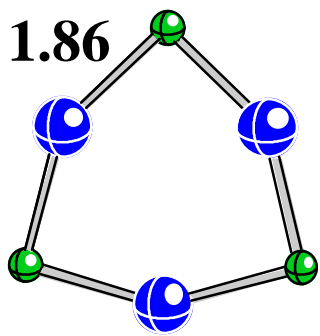

(e) 1.86  0μ_B

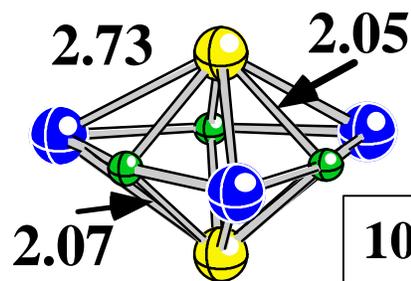

(f) 2.73 2.05 2.07  10μ_B

Fig. 4